\documentclass[letterpaper,titlepage,11pt]{article}

\pdfoutput=1

\usepackage{amssymb,amsmath,amsfonts}
\usepackage{epsfig}
\usepackage{ccaption}
\usepackage{graphicx}

\usepackage[
      colorlinks=true,
      linkcolor=blue,
      urlcolor=blue,
      filecolor=black,
      citecolor=red,
      pdfstartview=FitV,
      pdftitle={},
        pdfauthor={Roberto Emparan, Kentaro Tanabe},
        pdfsubject={},
        pdfkeywords={},
        pdfpagemode=None,
        bookmarksopen=true
      ]{hyperref}

\def\clock{{\count0=\time
           \divide\count0 60
           \ifnum\count0<10 0\fi\the\count0
           \multiply\count0 -60 \advance\count0 \time
           :\ifnum\count0<10 0\fi \the\count0
         }}
\newcommand{\timestamp}{{\small\vbox{\hbox{\tt\jobname.tex}
\hbox{\the\day/\the\month/\the\year, \clock}}}}

\newcommand{\ie}{{\it i.e.,\,}}
\newcommand{\eg}{{\it e.g.,\,}}

\newcommand{\lp}{\left(}
\newcommand{\rp}{\right)}

\newcommand{\mc}[1]{\mathcal{#1}}

\newcommand{\beq}{\begin{equation}}
\newcommand{\eeq}{\end{equation}}
\newcommand{\bea}{\begin{eqnarray}}
\newcommand{\eea}{\end{eqnarray}}
\newcommand{\beqa}{\begin{eqnarray}}
\newcommand{\eeqa}{\end{eqnarray}}

\newcommand{\Imm}{\mathrm{Im}\,}
\newcommand{\Ree}{\mathrm{Re}\,}

\begin{document}

\begin{titlepage}
\leftline{}
\vskip 2cm
\centerline{\LARGE \bf Universal quasinormal modes}
\medskip
\centerline{\LARGE \bf of large $D$ black holes 
} 
\vskip 1.6cm
\centerline{\bf Roberto Emparan$^{a,b}$, Kentaro Tanabe$^{b}$}
\vskip 0.5cm
\centerline{\sl $^{a}$Instituci\'o Catalana de Recerca i Estudis
Avan\c cats (ICREA)}
\centerline{\sl Passeig Llu\'{\i}s Companys 23, E-08010 Barcelona, Spain}
\smallskip
\centerline{\sl $^{b}$Departament de F{\'\i}sica Fonamental, Institut de
Ci\`encies del Cosmos,}
\centerline{\sl  Universitat de
Barcelona, Mart\'{\i} i Franqu\`es 1, E-08028 Barcelona, Spain}
\vskip 0.5cm
\centerline{\small\tt emparan@ub.edu,\, ktanabe@ffn.ub.es}

\vskip 1.6cm
\centerline{\bf Abstract} \vskip 0.2cm \noindent

We show that in the limit where the number of spacetime dimensions $D$ grows to infinity a very large class of black holes (including non-extremal, static, asymptotically flat ones, with any number of gauge-field charges, possibly coupled to dilatons) possess a universal set of quasinormal modes whose complex frequencies depend only on the horizon radius and no other black hole parameters. The damping ratio of these modes vanishes like $D^{-2/3}$, so they are almost normal modes, or `quasi-particle' excitations of the black hole. 
The structure responsible for the existence of these modes at large $D$ is also present very generally in other black holes.

\end{titlepage}
\pagestyle{empty}
\small
\normalsize
\newpage
\pagestyle{plain}
\setcounter{page}{1}




\paragraph{I.} If we probe a black hole by perturbing it away from equilibrium, its response ---the radiation it emits--- is dominated by the spectrum of its quasinormal modes. In many respects these are analogous to normal modes, but they have a dissipative part (imaginary frequency) due to the absorptive nature of the horizon. They can be regarded as the `free' (but damped) oscillations of the black hole spacetime, and so they provide a way of characterizing it. Black holes famously require very few parameters for their complete characterization, but their quasinormal modes are generically expected to carry the imprint of all of them. This is indeed borne out by the analytic methods known to approximate their calculation (for a review, see \cite{Berti:2009kk,Konoplya:2011qq}).  


Recently it has been argued that black holes and their dynamics simplify greatly in the limit in which the number $D$ of spacetime dimensions diverges \cite{Asnin:2007rw,Emparan:2013moa,Emparan:2013xia,Emparan:2013oza}. It is natural to ask whether this limit is useful in the calculation of quasinormal modes, and if so, what it reveals. We will see that at large $D$ an important part of the spectrum is not only very easy to compute, but also it is universally shared by many black holes, conveying only minimal information about the horizon radius $r_0$. If this property had been discovered through case-by-case numerical computation of quasinormal spectra for large values of $D$, it would probably have been regarded as a surprise. We will argue that it is a direct consequence of generic features of black holes at large $D$ discovered in \cite{Emparan:2013moa,Emparan:2013xia}.

More precisely, we shall show that for a large class of static, asymptotically flat, non-extremal black holes, there exist a number $\propto D^2$ of quasinormal modes whose complex frequencies at large $D$ are
%
\beq\label{uniom}
\omega_{(\ell,k)}\, r_0=\frac{D}2 +\ell -\lp \frac{e^{i\pi}}{2}\lp\frac{D}2 +\ell\rp\rp^{1/3}a_k\,,
\eeq
where $\ell$ is the angular momentum number, and $k=1,2,\dots$ is the `overtone' number with $-a_k$ being zeroes of the Airy function Ai. These are very well approximated by
\beq\label{akasy}
a_k\simeq \lp\frac{3\pi}{8}(4k-1)\rp^{2/3}\,.
\eeq

All the information about, \eg\ the gauge charges of the black hole, or their coupling to scalars (dilatons), has been effaced from \eqref{uniom}. It will be clear from our analysis that the structure responsible for this spectrum is also present in many other settings. 

The result \eqref{uniom} is valid for modes for which $\ell/D$ and $k$ are parametrically of order $D^0$ (\ie\ $\ell\ll D^2$, $k\ll D$). Then $\Ree\omega$ and $\Imm\omega$ are of order $D$ and $D^{1/3}$, respectively.
Among the entire set of quasinormal modes of large $D$ black holes, these
are not only most numerous, but they also have the smallest damping ratio $\Imm\omega/\Ree\omega\propto D^{-2/3}$. Their lifetime is very long on the time scale of their vibrational period, so they resonate most sharply, almost like a normal mode, to an external influence of the appropriate frequency.
Black holes have other quasinormal modes that do not conform to this spectrum, \eg\ with frequencies parametrically smaller, $\omega=O(D^0)$. These can be of considerable interest \cite{Hartnett:2013fba,toappear} but they do not seem to possess a similar degree of universality.


\paragraph{II.} 
Consider a static, spherically symmetric black hole with metric of the generic form
\beq\label{backgr}
ds^2= -h(r)A(r)B(r)dt^2+B(r)\frac{dr^2}{h(r)}+r^2 d\Omega_{D-2}^2\,,
\eeq
where
\beq
h=1-\lp\frac{r_0}{r}\rp^{D-3}
\eeq
fixes the position of the horizon at $r=r_0$. The functions $A(r)$, $B(r)$ are assumed to be non-zero and finite at all $r\geq r_0$ (hence we exclude extremal black holes). Asymptotic flatness requires that $A,B\to 1$ as $r\to\infty$. Beyond these basic assumptions, the class of solutions to which our argument below applies is very broad but not easily specified in full generality. A simple and very inclusive requirement is that when we take the large $D$ limit, 
the metric functions $A$ and $B$ approach $1$ exponentially fast in $D$ at any $r>r_0$.
This is indeed expected for any black hole that only supports massless fields outside its horizon (no massive hairs). We may then require that
at any given value of $r> r_0$, $A$ and $B$ behave like
\beq
A\to 1+ A_0(r)\lp\frac{r_0}{r}\rp^{D}+ \mc{O}\lp\frac{r_0}{r}\rp^{2D}\,,\qquad 
B\to 1+ B_0(r)\lp\frac{r_0}{r}\rp^{D}+ \mc{O}\lp\frac{r_0}{r}\rp^{2D}\,,
\eeq
where $A_0$ and $B_0$ remain finite and possibly non-zero in the limit. This condition is satisfied by, \eg\ dilatonic black holes, or black holes with charges coupled to $U(1)^n$ gauge theories (at least as long as the dilaton coupling, or $n$, remain finite as $D\to\infty$).

Linear perturbations of this black hole depend on the specific Lagrangian of the theory. However, among the gravitational perturbations, those that are tensors of $SO(D-1)$ do not couple to vector or scalar fields in the theory. The number of polarizations of tensor modes grows like $D^2$, much larger than vectors, $\sim D$, or scalars, $\sim D^0$. Assuming they are governed by Einstein's equations, the tensor modes satisfy the equation for a minimal massless scalar field $\Phi$ in the background spacetime \eqref{backgr} \cite{Kodama:2003jz}. Separating variables as $\Phi=r^{-(D-2)/2}\phi(r)e^{-i\omega t} Y^{(\ell)}_{D-2}(\Omega)$, this equation is
\beq
-\frac{d^2 \phi}{dr_*^2}+V(r_*)\phi =\omega^2\phi
\eeq
with
\beq\label{Veff}
V(r_*)=\frac{h A}{r^2}\left(
\ell(\ell+D-3)B -\frac{(D-2)^2}{4}h +\frac{(D-3)(D-2)}{2}+\frac{D-2}{4}h \frac{d\ln A}{d\ln r}
\right)
\eeq
and $dr_*=dr/(h\sqrt{A})$, so that $r_*\to -\infty$ at the horizon. For large $D$, $r_*$ approaches very closely the coordinate $r$ wherever $r>r_0(1+(\ln D)/D)$.

\begin{figure}[t]
\begin{center}
  	\includegraphics[width=.75\textwidth]{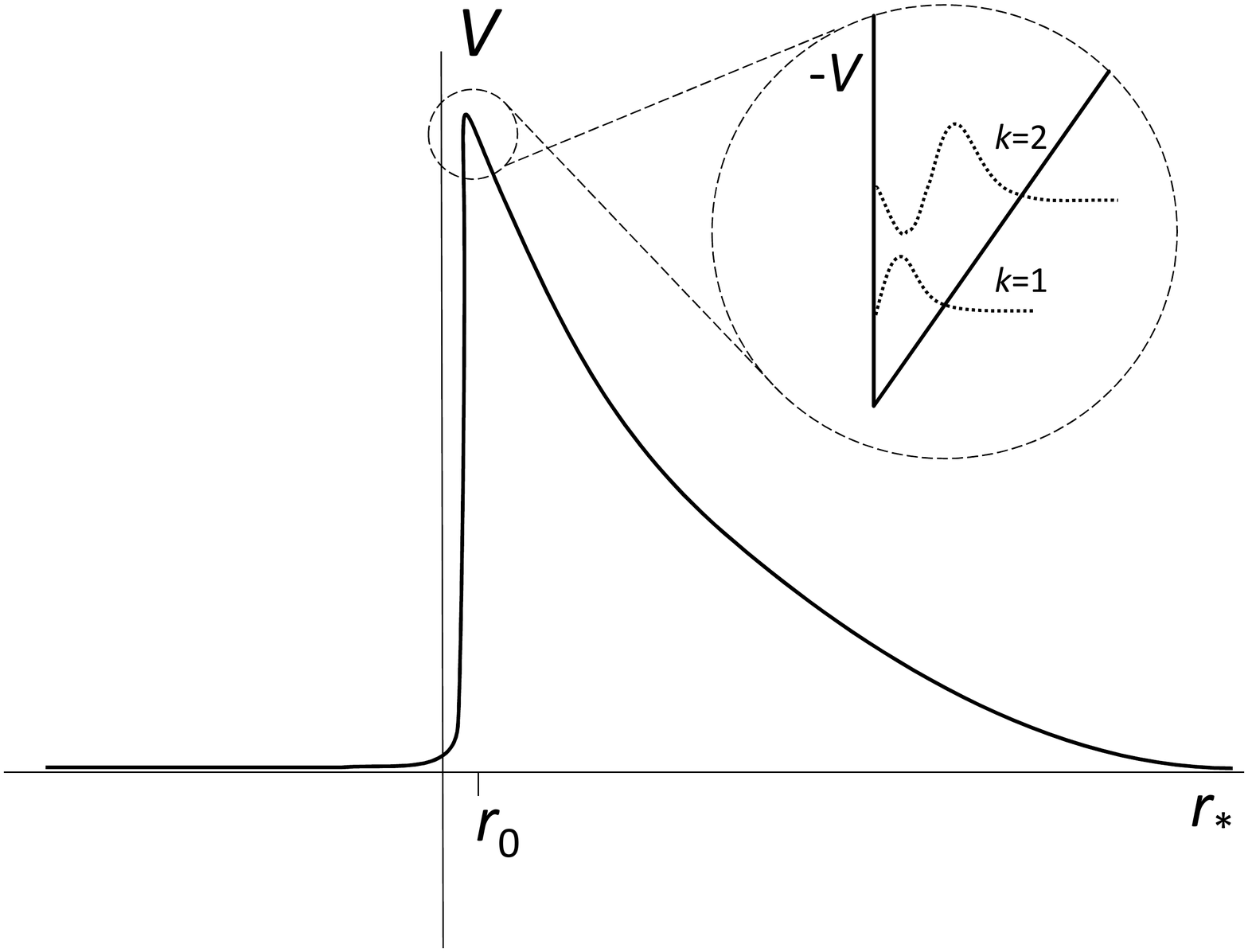}
\end{center}
\caption{\small Effective potential $V(r_*)$ at large $D$. When $D\to\infty$, $V$ is given by \eqref{niftyV} and the maximum becomes a sharp peak. Quasinormal modes correspond to bound states in the inverted potential $-V$ (inset). The lowest ones are the Airy-function bound states of a triangular well.}
\label{fig:potential}       
\end{figure}

The function $V$ describes a potential barrier that vanishes at infinity and at the horizon, and will reach a maximum at finite $r_*$. 
Now, under the assumptions above, when $D$ is very large, the potential can be written (up to subleading terms in $1/D$) in the form
\beq
V(r_*)= \frac{D^2\varpi_\ell^2}{r^2} h\lp 1+ C(r)\lp\frac{r_0}{r}\rp^D+ \mc{O}\lp\frac{r_0}{r}\rp^{2D}\rp\,,
\eeq
with 
\beq\label{omec}
\varpi_\ell\equiv\frac12 +\frac{\ell}{D}\,,
\eeq
and where the function $C(r)$ remains finite at any $r>r_0$ as $D\to\infty$. It is now easy to see that this kind of potential
has a maximum at 
\beq
r_*^\mathrm{max}=r_0\lp 1+\frac{\ln a D}{D}\lp 1+\mc{O}(1/D)\rp\rp\,,
\eeq
(with constant $a$) and therefore $r_*^\mathrm{max}\to r_0$ as $D\to\infty$, whenever $C(r_0)<1$. The latter condition holds for, \eg\ dilatonic black holes, but when it is not satisfied the maximum is expected at $r_*^\mathrm{max}=r_0(1+b/D)$ (with constant $b>0$) so again $r_*^\mathrm{max}\to r_0$. The potential around this maximum is very simple: to its left, $r_*<r_*^\mathrm{max}$, we have $h\sim e^{D (r_*-r_*^\mathrm{max})}$ so the potential drops to zero exponentially. To its right, $r_*>r_*^\mathrm{max}$, we have that $(r_0/r)^D$ is exponentially small in $D$, so in this region $V$ becomes the radial potential of the scalar field in Minkowski space. Then, when $D\to\infty$  the potential becomes simply (see fig.~\ref{fig:potential})
\beq\label{niftyV}
V(r_*)\to \frac{D^2 \varpi_\ell^2}{r_*^2}\, \Theta(r_*-r_0)\,.
\eeq

The upshot of this discussion should be plain now: very generally, as a consequence of the expected behavior of the metric functions $A$ and $B$ as $D\to\infty$, we find a flat space radial potential cutoff at $r_*=r_0$, with $V_\mathrm{max}\propto D^2$. As will be clear presently, we are only interested in the potential close to this maximum. In some cases there may be other subleading extrema to the left of this maximum, with $V\sim O(D^0)$, which we shall not consider. 

We now seek the quasinormal modes of the potential \eqref{niftyV}, \ie\ solutions satisfying $\phi\sim e^{\pm i\omega r_*}$ at $r_*\to \pm\infty$. Upon continuation $r_*\to -i r_*$, $\varpi_\ell\to i\varpi_\ell$, these can be equivalently obtained as the bound states in the inverted potential $-V$, satisfying $\phi\sim e^{\mp \omega r_*}$ at $r_*\to \pm\infty$ \cite{Ferrari:1984zz}. 
We are interested in the least damped modes, which correspond to the lowest bound states in $-V$. At large $D$ these are very strongly localized near the minimum, which can be approximated as a triangular well. There, defining $x=(r_*-r_0)/r_0$, the wave equation for $x\geq 0$ is
\beq\label{tipeq}
-\frac1{D^2}\frac{d^2\phi}{dx^2}+ 2 \varpi_\ell^2\, x\, \phi(x) =\lp \varpi_\ell^2-\frac{\omega^2 r_0^2}{D^2}\rp\phi(x)\,.
\eeq 
Up to normalization, the bound state wavefunctions are
\beq\label{phisol}
\phi=\mathrm{Ai}\left[\lp\frac{D}{2\varpi_\ell^2}\rp^{2/3}\lp 2\varpi_\ell^2 x-\varpi_\ell^2+\frac{\omega^2 r_0^2}{D^2}\rp
\right]\,.
\eeq
Since the depth of the well diverges $\propto -D^2$, we must impose $\phi(x=0)=0$. This quantizes the values of $\omega$ in terms of the Airy zeroes $\mathrm{Ai}(-a_k)=0$. Then, after undoing the analytic continuation, we get the spectrum \eqref{uniom}.
The asymptotic formula
\beq
\mathrm{Ai}(-z)=\frac1{\sqrt{\pi}\, z^{1/4}}\cos\lp\frac{2z^{3/2}}{3}-\frac{\pi}4\rp\lp 1+O\lp z^{-3/2}\rp\rp
\eeq
yields the approximation \eqref{akasy}, which estimates the first Airy zero $a_1=2.338$ to better than $1\%$ accuracy.

\paragraph{III.} 
This universal spectrum is easy to understand: when $D\to\infty$, the geometry outside the horizon becomes Minkowski space, and nothing specific remains of the black hole geometry other than an abrupt cutoff at $r_0$. We find the physics of scalar oscillations on a hole of radius $r_0$, which absorbs perfectly all waves with frequencies larger than $D\varpi_\ell/r_0$ and resonates quasi-normally with the modes \eqref{uniom} just below this threshold.

Ref.~\cite{Emparan:2013moa} argued that these geometries of holes cut out in Minkowski or (A)dS space appear very generally in the limit $D\to\infty$ of black hole spacetimes, including black holes with charge, rotation, or cosmological constant. This structure gives effective potentials for wave propagation with maxima that become triangular-shaped when $D\to\infty$. So we expect quasinormal spectra of the form \eqref{uniom} in many other instances ---possibly replacing $\ell$ with eigenvalues of the Laplacian in the transverse space, or changing the slope of the triangular well depending on the background geometry--- but still carrying minimal information about the black hole size and none of its other parameters. For Schwarzschild-AdS black holes, the peak is present only for small black holes, more precisely, $r_0<L \arctan(1/\sqrt{2})$ for $\ell<O(D)$ (small AdS black holes also have another set of modes with lower damping ratio, $\Imm\omega/\Ree\omega\sim (r_0/L)^{D+2\ell}$).
The quasinormal modes of rotating black holes are somewhat more complicated, but we expect that these features still play a role.

The simplicity of the argument allows to extend the analysis to gravitational vector and scalar perturbations. The radial potentials for these perturbations of Schwarzschild black holes are more complicated than \eqref{Veff} \cite{Kodama:2003jz}, but in the leading large $D$ limit they possess the same peak as in \eqref{tipeq}, so we recover the universal modes. These potentials have further structure that is responsible for other quasinormal modes not captured by this analysis, with $\Imm\omega$ of order $D^0$, so they are longer-lived than the universal set, but they are also broader, with $\Imm\omega/\Ree\omega\sim D^0$.


Our computation is largely independent of the specifics of the near-horizon geometry, but it makes an implicit assumption of genericity. It is easy to see that the generic behavior around the peak region (where $1/D\ll x\ll 1$) of a near-horizon solution $\phi(x)$ is that $D \phi/\phi'|_{x=0}=O(D^0)$, which for $\eqref{phisol}$ indeed requires that $\phi(x=0)\to 0$. The boundary conditions imposed by non-extremal horizons do indeed generically lead to this behavior, but this may fail for extremal black holes (possibly also for black holes exponentially close (in $D$) to extremality), for which a detailed matched asymptotic construction of the near-horizon and far-zone solutions is needed --- similar to the conclusion in \cite{Emparan:2013oza}. We expect to provide more details on this issue elsewhere.

Let us now comment on eq.~\eqref{uniom} as an approximation to the quasinormal spectrum at finite $D$. First, observe that it may be appropriate to replace $D\to D-3$ in \eqref{uniom}, since it is $D-3$ that \eg\  controls the fall-off of the gravitational field. Second, an often useful method to compute quasinormal spectra is a WKB approximation in which the radial potential $V$ is expanded near the maximum to quadratic order, and thus replaced by an inverted parabola \cite{schutz}. For the $D$-dimensional Schwarzschild black hole this method \cite{Konoplya:2003ii,Berti:2003si} yields, at large $D$, the same leading value $\Ree\omega=D\varpi_\ell/r_0=\sqrt{V_\mathrm{max}}$, but it would predict that $\Imm\omega \sim D^{1/2}$. This is incorrect: the parabola is a bad approximation when the maximum becomes very pointy. Still, this suggests that, as a function of $D$, $\Imm\omega$ could change from $\sim D^{1/2}$ at moderate $D$, to $\sim D^{1/3}$ at very large $D$. This change may not be easy to detect numerically. A naive estimate suggests that the crossover may occur at rather large values of $D$ (possibly $>20$), so, without further analysis, it is unclear how accurately  eq.~\eqref{uniom} gives the value of $\Imm\omega$ at moderately large $D$.

To conclude, we have identified a set of quasinormal black hole oscillations with remarkable properties: not only are they very simply identified and universally present for large classes of black holes, but they are also highly degenerate and very sharp (long-lived in their own time scale $r_0/D$: the `string scale' of \cite{Emparan:2013xia}). All these may be indications of an important role of these modes for a further understanding, possibly microscopic, of black holes within the large $D$ expansion.

\section*{Acknowledgments}

We thank Vitor Cardoso for useful comments. Work partially supported by MEC FPA2010-20807-C02-02, AGAUR 2009-SGR-168 and CPAN CSD2007-00042 Consolider-Ingenio 2010.
KT is supported by a grant for research abroad by JSPS.


\end{document}